# Large Teams Consistently Develop Science

Yiling Lin[1], Lulin Yang[1*] and Lingfei Wu[1*]

[1] School of Computing and Information, The University of Pittsburgh, 135 N Bellefield Ave, Pittsburgh, PA 15213

[*] Corresponding author. E-mail: liw105@pitt.edu (L.W)

## Abstract

Scientific progress depends on both developing established knowledge and disrupting it to make way for new ideas. Using the disruption index (D), our previous work showed that large teams tend to develop science, while small teams are more likely to disrupt it. Recent work challenges this finding, arguing that D mechanically depends on reference-list length. Using an alternative measure (oneD), it reaches the opposite conclusion: large teams are more disruptive. Combining Nobel Prize case studies, mathematical analysis, and large-scale publication data, we show that oneD misclassifies developmental contributions as disruptive and introduces the very sensitivity to reference-list length that it seeks to eliminate. The original measure, D, however, shows no evidence of the claimed systematic dependence on reference-list length. More fundamentally, even after controlling for reference-list length, larger teams remain more likely to produce strongly developmental work. Our findings reaffirm that large teams consistently develop science and offer new insights into the measurement of scientific disruption.

### A Nobel Discovery, Two Disruption Measures

The discovery of Bose–Einstein condensation in dilute atomic gases was recognized by the 2001 Nobel Prize in Physics. In 1995, Anderson et al. first demonstrated condensation in rubidium atoms. Soon after, Davis et al. achieved condensation in sodium atoms, producing a condensate more than a hundred times larger. Rather than replacing the earlier discovery, this work extended and consolidated it, establishing Bose–Einstein condensation as an experimentally accessible phenomenon.

Yet the Davis paper had very different relationships with its individual references (Fig. 1). Its pairwise disruption score with Anderson et al., its most-cited reference (over 5,000 citations), was −0.59, indicating a strongly developmental relationship. Its relationship with Bradley et al.'s parallel discovery in lithium (over 2,000 citations) was nearly neutral (−0.04). In contrast, its pairwise scores with earlier experimental approaches were strongly positive, including Petrich et al.'s magnetic trap (about 350 citations; +0.89) and Adams et al.'s optical trap (about 150 citations; +0.96). Thus, the same Nobel Prize-winning discovery consolidated a foundational advance while departing from several earlier experimental approaches.

The original disruption index assigns Davis et al. a score of −0.55, consistent with its developmental contribution. Yet the recently proposed one-reference disruption index (oneD; Huang and Xie, 2026), which averages pairwise scores from randomly sampled references, yields +0.61, characterizing the same paper as disruptive. This example illustrates how oneD can misclassify a major developmental advance as disruptive, raising questions about its ability to distinguish between these two forms of scientific progress.

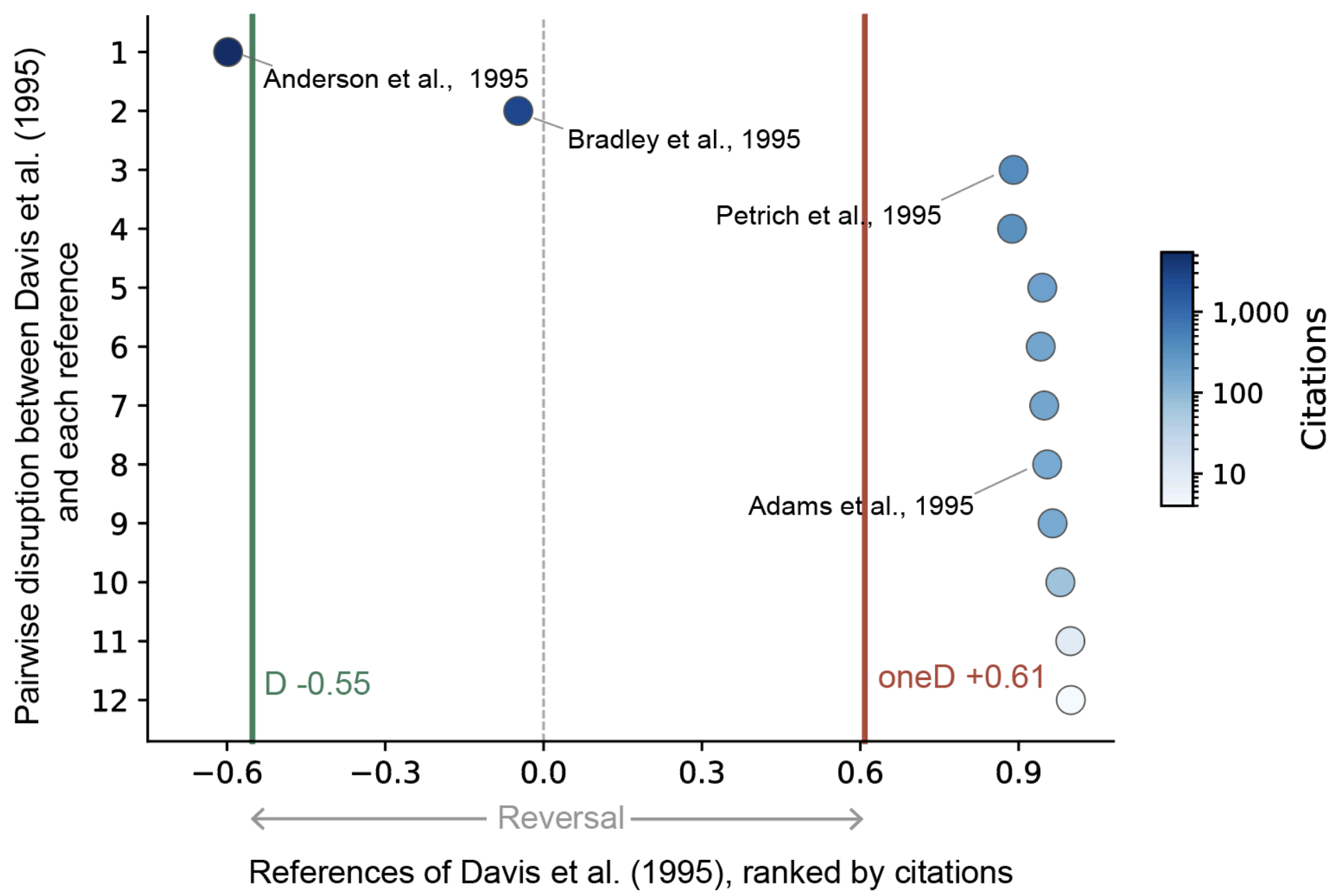

**Figure 1. Pairwise disruption scores for a Nobel Prize-winning discovery.** Davis et al. (1995) extended Anderson et al.'s demonstration of Bose–Einstein condensation in rubidium to sodium, producing a larger condensate. Each circle represents one of the 12 journal-article references of Davis et al. identified in OpenAlex, ranked by citation count on the y-axis. The x-axis shows pairwise disruption scores (d), with circle color indicating citation count on a logarithmic scale. The green line indicates the original disruption index ($D = -0.55$), and the red line indicates oneD (+0.61), estimated by averaging pairwise scores from 30 draws with replacement from the paper's references.

## Two Kinds of Scientific Advances

Is the Davis paper an exceptional case, or does this divergence extend to other major scientific discoveries? To investigate, we examine 511 Nobel Prize-winning papers spanning more than a century, drawn from the dataset of Li et al. (2019) and matched to OpenAlex, for which both D and oneD can be computed. Under D, 107 papers (21%) receive negative scores, indicating contributions that develop and consolidate existing knowledge rather than displace it. Yet under oneD, 106 of these 107 papers receive positive scores, virtually eliminating the developmental category.

The difference becomes more apparent when Nobel-winning papers are compared with ordinary scientific publications. Under D, Nobel papers are disproportionately represented at both ends of the disruption spectrum, producing a U-shaped pattern: they are three to six times more prevalent among strongly developmental papers and more than ten times as prevalent among highly disruptive papers, while being substantially underrepresented near zero. This pattern highlights two distinct properties of D: its magnitude helps distinguish major scientific advances from ordinary work, while its sign distinguishes advances that develop existing knowledge from those that disrupt it. Under oneD, the developmental side of this distribution largely disappears, with Nobel papers concentrated at high positive scores (Fig. 2A-B).

Huang and Xie (2026) reported that oneD outperforms D in identifying Nobel Prize-winning papers. To examine this claim, we reproduce their comparison and introduce absolute disruption (absD = |D|), which preserves the magnitude of D while discarding its sign. We evaluate all three measures using receiver operating characteristic (ROC) curves, which compare the true positive rate against the false positive rate across classification thresholds. The area under each curve (AUC) summarizes its ability to distinguish Nobel-winning papers from ordinary publications (Fig. 2C). Consistent with Huang and Xie (2026), oneD achieves a higher AUC than D (0.97 versus 0.78). However, taking the absolute value of D also raises its AUC to 0.89.

This comparison reveals why oneD's apparent advantage requires careful interpretation. Part of its advantage over D can be reproduced simply by ignoring the sign of D and focusing on its magnitude. A higher AUC does not necessarily indicate a more informative measure of scientific contribution. It rewards the ability to distinguish major scientific advances from ordinary work, but not the ability to distinguish between two kinds of advances: those that develop existing knowledge and those that disrupt it.

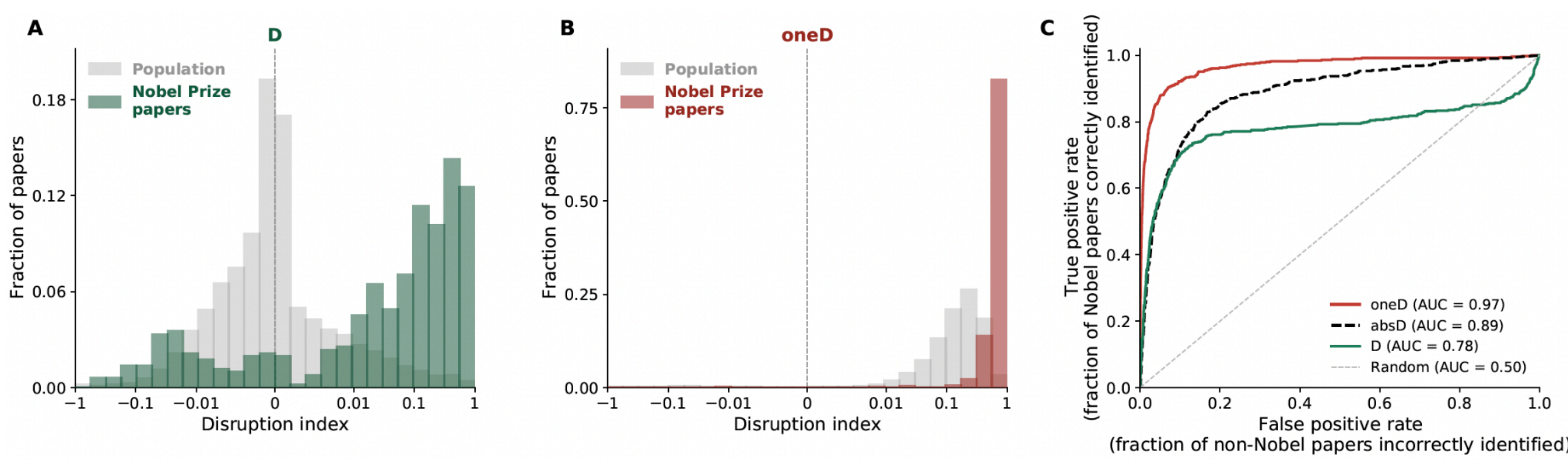

**Figure 2. Disruption measures and Nobel Prize-winning papers.** We analyzed 511 Nobel Prize-winning papers from Li et al. (2019) and 50,000 randomly sampled papers from OpenAlex. (A) Distribution of the original disruption index (D) for Nobel papers (green) and the comparison population (gray). (B) Distribution of oneD for Nobel papers (red) and the comparison population (gray). (C) Receiver operating characteristic (ROC) curves for distinguishing Nobel papers from the comparison population using D (green), its absolute value absD (black dashed), and oneD (red). The corresponding areas under the curve (AUCs) are 0.78, 0.89, and 0.97, respectively.

## The oneD Paradox

The preceding analysis reveals a striking tendency of oneD to classify developmental contributions as disruptive. To understand why, we examine two properties of scientific citations: a mathematical asymmetry in pairwise disruption scores and a highly unequal distribution of citation impact across references. Together, these properties help explain how oneD can produce systematically positive scores even for developmental discoveries.

The first property follows directly from the definition of pairwise disruption. Consider a focal paper cited by $N$ subsequent papers. For each reference $r$, let $a_r$ denote the number of papers citing both the focal paper and $r$, and $b_r$ the number citing $r$ but not the focal paper. The pairwise disruption score is

$$d_r = \frac{N-2a_r}{N+b_r} \quad (1)$$

A negative score requires $a_r > N/2$. Since $a_r$ cannot exceed the reference's citation count $C_r = a_r + b_r$, we obtain the necessary condition

$$d_r < 0 \quad \Longrightarrow \quad C_r > \frac{N}{2} \tag{2}$$

Highly cited references can therefore produce either positive or negative scores, depending on their citation relationships with the focal paper. In contrast, references with fewer than half as many citations as the focal paper cannot produce negative scores. Thus, the sign of pairwise disruption is asymmetrically constrained by reference citation impact, independently of how reference citations are distributed.

The second property concerns the distribution of citation impact. Previous research has established that scientific citations are highly unequal, with a small fraction of publications receiving a disproportionate share of citations (Barabási and Albert, 1999; Wang et al., 2013). Our recent work extends this pattern to the references within individual papers, showing that their citation counts follow a shifted Zipf law (Lin et al., 2026):

$$C_r = A(\beta + r)^{-\alpha} \tag{3}$$

where $r$ denotes the rank of a reference ordered by decreasing citation count, $\alpha$ characterizes the inequality of the distribution, and $\beta$ and $A$ are fitted parameters. As reference rank increases, citation counts decline sharply, leaving a small number of influential predecessors and a long tail of less-cited references.

These two properties interact. Because $a_r \leq C_r$ and $b_r \geq 0$, Eq. (1) implies

$$d_r \geq 1 - \frac{2C_r}{N} = 1 - \frac{2A}{N(\beta+r)^{\alpha}} \tag{4}$$

Under a shifted Zipf law, this lower bound increases toward +1 as reference rank increases, holding the focal paper's citation count and the distribution parameters fixed. Consequently, references in the long tail become increasingly constrained in their ability to produce negative pairwise disruption scores. The references most capable of producing strongly negative scores are concentrated near the top of the citation ranking, while sufficiently low-impact references necessarily have pairwise scores close to +1.

The paradox arises from how oneD aggregates these relationships. By uniformly sampling references and averaging their pairwise disruption scores, oneD assigns equal weight to references with highly unequal citation influence. For a focal paper with $R$ references, its expected value is

$$E[oneD] = \frac{1}{R}\sum_{r=1}^{R} d_r \tag{5}$$

Under the shifted Zipf model, Eq. (4) implies

$$E[oneD] \geq 1 - \frac{2A}{NR}\sum_{r=1}^{R} \frac{1}{(\beta+r)^{\alpha}} \tag{6}$$

For $\alpha > 0$ and fixed $N, b,$ and $c$, the average on the right tends to zero as $R$ increases. Since pairwise disruption scores cannot exceed +1, the expected oneD score consequently approaches +1. Thus, the numerous references in the long tail can dominate the average, shifting oneD toward positive values even when a paper makes a substantial developmental contribution to influential predecessors.

This mechanism predicts that oneD may increase systematically with reference-list length when longer lists extend the low-impact tail. Because larger teams also tend to cite more references (Huang and Xie, 2026), this sensitivity may distort the observed relationship between team size and disruption. We examine these predictions using large-scale publication data.

**Team Size, Reference Length, and Disruption**

We analyze 9,394,828 journal articles from OpenAlex, published between 1900 and 2015, with citations tracked through 2025 to allow sufficient time for disruption scores to stabilize. We examine how D and oneD vary with reference-list length while holding team size constant, and whether larger teams remain more likely to produce strongly developmental work (bottom 5% D) after controlling for reference-list length.

Among papers with the same team size, oneD increases systematically with reference-list length, consistent with our mathematical analysis (Fig. 3A). In contrast, D shows no consistent trend with reference-list length, contrary to Huang and Xie's (2026) argument.

We then examine whether larger teams remain more likely to produce developmental work when reference-list length is held constant (Fig. 3B). Among papers with the same number of references, multi-author teams are consistently more likely than single authors to fall in the bottom 5% of D, indicating strong developmental contributions. This difference is particularly pronounced for teams of ten or more authors.

Together, these findings show that the original relationship between team size and disruption persists after controlling for reference-list length. They also suggest that the opposite relationship reported using oneD may be driven, at least in part, by its sensitivity to reference-list length. Thus, larger teams remain more likely to produce strongly developmental work, independent of reference-list length.

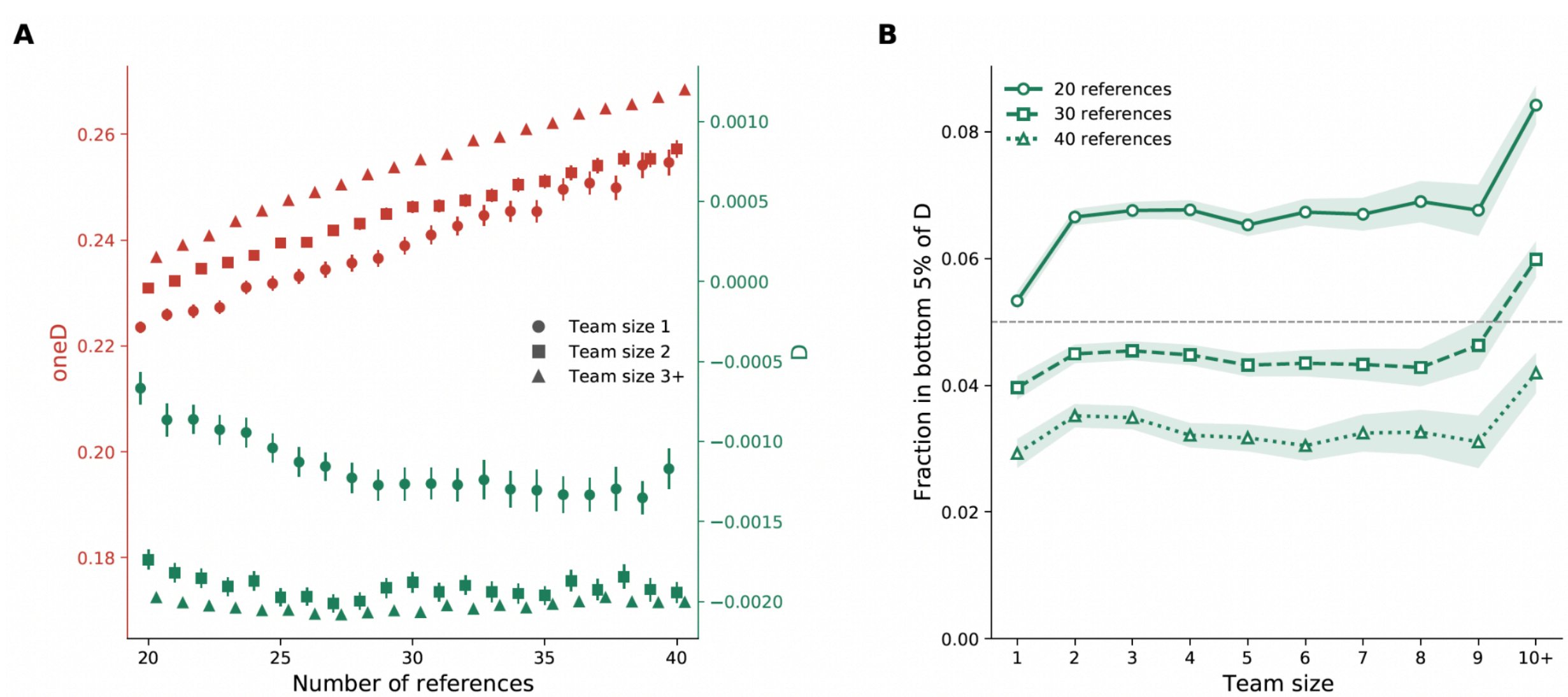

**Figure 3. Disruption measures across reference-list lengths and team sizes.** We analyzed 9,394,828 journal articles from OpenAlex, published between 1900 and 2015, with citations tracked through 2025 to allow sufficient time for disruption scores to stabilize. **(A)** Average disruption scores measured by oneD (red, left axis) and D (green, right axis) across reference-list lengths, separately for papers with one, two, and three or more authors. **(B)** Fraction of papers in the bottom 5% of D by team size, separately for papers with 20, 30, and 40 references. Error bars and shaded bands represent 95% confidence intervals estimated by bootstrapping.

## Conclusion

Our analysis shows that oneD's equal averaging across references can misclassify developmental contributions as disruptive under highly unequal citation distributions. In contrast, the original disruption measure, D, shows no systematic dependence on reference-list length. Crucially, even after controlling for reference-list length, large teams consistently develop science.

This finding matters because development is no less important than disruption. As Nobel Prize-winning discoveries illustrate, both highly disruptive work and highly developmental work represent major scientific advances. Understanding scientific progress therefore requires measures that capture not only the magnitude of a contribution but also its direction: whether new knowledge builds upon what came before or displaces it.